\documentclass[12pt]{article}
\usepackage{amsbsy}
\usepackage{amsfonts}
\usepackage{amsmath}
\usepackage{amssymb}
\usepackage{apacite}
\usepackage{setspace}
\usepackage{graphicx}
\usepackage{bm,multicol}
\usepackage[english]{babel}
\usepackage{natbib}
\usepackage[T1]{fontenc}
\usepackage[utf8]{inputenc}
\usepackage{authblk}
\usepackage{xcolor}

\newtheorem{lemma}{Lemma}

\newtheorem{remark}{Remark}

\newcommand{\newc}{\newcommand}
\newc{\N}{\mbox{N}}
\newc{\1}{\bf{1}}

\begin{document}
\title{A Generalization of the Savage-Dickey Density Ratio for Testing Equality and Order Constrained Hypotheses}

\author{J. Mulder, E.-J. Wagenmakers \& M. Marsman}

\date{}

\maketitle

\begin{abstract}
The Savage-Dickey density ratio is a specific expression of the Bayes factor when testing a precise (equality constrained) hypothesis against an unrestricted alternative. The expression greatly simplifies the computation of the Bayes factor at the cost of assuming a specific form of the prior under the precise hypothesis as a function of the unrestricted prior. A generalization was proposed by Verdinelli and Wasserman (1995) such that the priors can be freely specified under both hypotheses while keeping the computational advantage. This paper presents an extension of this generalization when the hypothesis has equality as well as order constraints on the parameters of interest. The methodology is used for a constrained multivariate $t$ test using the JZS Bayes factor and a constrained hypothesis test under the multinomial model.
\end{abstract}


\section{Introduction}
The Savage-Dickey density ratio \citep{Dickey:1971} is a special expression of the Bayes factor, the Bayesian measure of statistical evidence between two statistical hypotheses in light of the observed data \citep{Jeffreys,Kass:1995}. The Savage-Dickey density ratio is relatively easy to compute from Markov chain Monte Carlo (MCMC) output without requiring the marginal likelihoods under the hypotheses. Consider a test of a normal mean $\theta$ with unknown variance $\sigma^2$, $H_c:\theta=0$ versus $H_u:\theta\in\mathbb{R}$, with independent observations $y_i\sim N(\theta,\sigma^2)$, for $i=1,\ldots,n$. The indices `$c$' and `$u$' refer to a constrained hypothesis and an unconstrained hypothesis\footnote{The test can equivalently be formulated as a test of $H_c:\theta=0$ versus $H_u:\theta\not = 0$ as $\theta=0$ has zero probability under $H_u$ when using a continuous prior for $\theta$. The formulation $H_u:\theta\in\mathbb{R}$ is used however to make it explicit that the constrained hypothesis $H_c$ is nested in the unconstrained hypothesis $H_u$.}. Denote the priors for the unknown parameters under $H_c$ and $H_u$ by $\pi_c(\sigma^2)$ and $\pi_u(\theta,\sigma^2)$, respectively, which reflect which values for the parameters are likely before observing the data. Under $H_u$ we consider a unit information prior $\pi_u(\theta|\sigma^2)=N(0,\sigma^2)$ and a conjugate inverse gamma prior for the nuisance parameter, say, $\pi_u(\sigma^2)=IG(\frac{1}{2},\frac{1}{2})$ \citep[the exact choice of the hyperparameters does not qualitatively affect the argument; see also][for example]{Verdinelli:1995}. The marginal prior for $\theta$ under $H_u$ then follows a Cauchy distribution (equivalent to a Student $t$ distribution with 1 degree of freedom) centered at $\theta=0$ with a scale parameter of 1. The marginal posterior for $\theta$ under $H_u$, $\pi_u(\theta|\textbf{y})$, also has a Student $t$ distribution. When the prior for the nuisance parameter $\sigma^2$ under $H_c$ equals the conditional prior for $\sigma^2$ under $H_u$ given the restriction under $H_c$, i.e., $\pi_c(\sigma^2)=\pi_u(\sigma^2|\theta=0)$, the Bayes factor for $H_c$ against $H_u$ can then be written as the Savage-Dickey density ratio: the ratio of the unconstrained posterior and unconstrained prior density evaluated at the constrained null value under $H_c$ \citep{Dickey:1971}, i.e., 
\[
B_{cu} = \frac{p_c(\textbf{y})}{p_u(\textbf{y})}=\frac{\int p(\textbf{y}|0,\sigma^2)\pi_1(\sigma^2) d\sigma^2}{\iint p(\textbf{y}|\theta,\sigma^2)\pi_u(\theta,\sigma^2)d\theta d\sigma^2}=
\frac{\pi_u(\theta=0|\textbf{y})}{\pi_u(\theta=0)},
\]
where $p(\textbf{y}|\theta,\sigma^2)$ denotes the likelihood of the data given the normal mean $\theta$ and variance $\sigma^2$, and $p_c(\textbf{y}$ and $p_u(\textbf{y})$ denote the marginal likelihoods under $H_c$ and $H_u$, respectively. For the current problem we would thus need to divide the posterior $t$ distribution of $\theta$ under $H_u$ evaluated at $\theta=0$ by the prior Cauchy distribution at $\theta=0$, which both have analytic expressions. Note, of course, that the same expression would be obtained by deriving the marginal likelihoods which also have analytic expressions in this scenario. For more complex statistical models with more nuisance parameters, for which the marginal likelihoods would not have analytic expressions, the Savage-Dickey density ratio is particularly useful as we only need to compute the ratio of the unconstrained posterior and the unconstrained prior evaluated at the constrained null value, which are generally easy to obtain, e.g., using MCMC output.

Despite its computational convenience, a limitation of the Savage-Dickey density ratio is that it only holds for a specific form of the prior for the nuisance parameters under the restricted model which is completely determined by the prior under the unrestricted model. This imposed prior under the restricted model may not always have a desirable interpretation. For example, in order for the Savage-Dickey ratio to hold in the above example, the prior for the population variance under $H_c$ equals $\pi_c(\sigma^2)=\pi_u(\sigma^2|\theta=0)=IG(1,\frac{1}{2})$. This prior under $H_c$ is more concentrated around smaller values for $\sigma^2$ than under $H_u$ as can be seen from the prior modes for $\sigma^2$ under $H_c$ and $H_u$ which are $\frac{1}{4}$ and $\frac{1}{3}$, respectively. This is contradictory however because the sample variance for $\sigma^2$ will always be smaller under $H_u$ when the mean $\theta$ is unrestricted. Therefore the Savage-Dickey density ratio should be used with care. For discussions on the Savage-Dickey density ratio, see \cite{MarinRobert:2010} and \cite{Heck:2020}. For discussions on priors for the nuisance parameters, see \cite{ConsonniVeronese2008}.

To retain the computational convenience of the Savage-Dickey density ratio, while allowing researchers to freely specify the prior for the nuisance parameters under the restricted model, \cite{Verdinelli:1995} proposed a generalization. In a multivariate setting when testing a vector of key parameters $\bm\theta$, i.e., $H_c:\bm\theta=\textbf{r}$, where $\textbf{r}$ is a vector of constants, against an unconstrained alternative, $H_u:\bm\theta$ unconstrained, with nuisance parameters $\bm\phi$, where the priors under $H_c$ and $H_u$ are denoted by $\pi_c(\bm\phi)$ and $\pi_u(\bm\theta,\bm\phi)$, respectively, the multivariate generalized Savage-Dickey density ratio is given by
\begin{equation}
B_{1u} = \frac{\pi_u(\bm\theta=\textbf{r}|\textbf{y})}{\pi_u(\bm\theta=\textbf{r})}
\times
\mathbb{E}\left\{ \frac{\pi_{c}(\bm\phi)}{\pi_{u}(\bm\phi|\bm\theta=\textbf{r})}
\right\},
\label{SD2}
\end{equation}
where the expectation is taken over the conditional posterior under the unconstrained model, $\pi_{u}(\bm\phi|\bm\theta=\textbf{r},\textbf{y})$. As can be seen, the generalization is equal to the original Savage Dickey density ratio (the first factor on the right hand side of \eqref{SD2}) multiplied with a correction factor based on the ratio of the freely chosen prior for the nuisance parameters, $\pi_c(\bm\phi)$, and the imposed prior for the nuisance parameters under the Savage-Dickey density ratio, $\pi_u(\bm\phi|\bm\theta=\textbf{r})$. In the above example, one might want to use the same marginal prior for the nuisance parameter under $H_c$ as under $H_u$, i.e., $\pi_c(\sigma^2)=IG(\frac{1}{2},\frac{1}{2})$.

The generalization in \eqref{SD2} was not derived when the constrained hypothesis contains order (or one-sided) constraints in addition to equality constraints, say, $H_c:\bm\theta_e=\textbf{r}_e~\&~\bm\theta_o>\textbf{r}_o$. Scientific theories however are very often formulated with combinations of equality and order constraints \citep{Hoijtink:2011}. In repeated measures studies for instance, theory may suggest a specific ordering of the measurement means \citep{DeJong:2017} or measurement variances \citep{BoingMessing2020}, in a regression model theory may suggest that a certain set of predictor variables have zero effects, while other variables are expected to have a positive or a negative effects \citep{MulderOlsson:2019}, or order constraints may be formulated on regression effects \citep{Haaf} or intraclass correlations \citep{MulderFox:2019} in multilevel models. The goal of the current paper is therefore to show the generalization of the Savage-Dickey density ratio in \eqref{SD2} for a constrained hypothesis with equality and order constraints on certain key parameters. This is shown in Section 2, where the generalization is related to existing special cases of the Bayes factor. Section 3 presents two applications of Bayesian constrained hypothesis testing under two statistical models: A multivariate Bayesian $t$ test for standardized effects under the multivariate normal model using a novel extension of the JZS Bayes factor \citep{Rouder:2009}, and a constrained hypothesis test on the cell probabilities under a multinomial model. The paper ends with some short concluding remarks in Section 4.

\section{Extending the Savage-Dickey density ratio}
Lemma \ref{lemmaSD} presents our main result.

\begin{lemma}\label{lemmaSD}
Consider a constrained statistical model, $H_c$, where the parameters $\bm\theta_e$ are fixed with equality constraints, i.e., $\bm\theta_e=\textbf{r}_e$, and order (or one-sided) constraints are formulated on the parameters $\bm\theta_o$, i.e., $\bm\theta_o>\textbf{r}_o$, with (unconstrained) nuisance parameters $\bm\phi$, and an alternative unconstrained model $H_u$, where $(\bm\theta_e,\bm\theta_o,\bm\phi)$ are unrestricted. If we denote the priors under $H_c$ and $H_u$ according to $\pi_c(\bm\theta_o,\bm\phi)$ and $\pi_u(\bm\theta_e,\bm\theta_o,\bm\phi)$, respectively, then the Bayes factor of model $H_c$ against model $H_u$ given a data set $\textbf{y}$ can be expressed as
\begin{equation}
B_{cu}=\frac{\pi_u(\bm\theta_e=\textbf{r}_e|\textbf{y})}{\pi_u(\bm\theta_e=\textbf{r}_e)\text{Pr}_{c^*}(\bm\theta_o>\textbf{r}_o)}\times
\mathbb{E}\left\{ \frac{\pi_{c^*}(\bm\theta_o,\bm\phi)}{\pi_{u}(\bm\theta_o,\bm\phi|\bm\theta_e=\textbf{r}_e)}
1_{\{\bm\theta_o>\textbf{r}_o\}}(\bm\theta_o)
\right\},
\label{SD3}
\end{equation}
where the expectation is taken over the conditional posterior of $(\bm\theta_o,\bm\phi)$ given $\bm\theta_e=\textbf{r}_e$ under $H_u$, i.e., $\pi_{u}(\bm\theta_o,\bm\phi|\textbf{y},\bm\theta_e=\textbf{r}_e)$, and $\pi_{c^*}(\bm\theta_o,\bm\phi)$ denotes the ``completed'' prior under the completed constrained hypothesis where the one-sided constraints are omitted, i.e., $H_{c^*}:\bm\theta_e=\textbf{r}_e$, such that $\pi_{c}(\bm\theta_o,\bm\phi)=\text{Pr}_{c^*}(\bm\theta_o>\textbf{r}_o)^{-1}\pi_{c^*}(\bm\theta_o,\bm\phi)1_{\{\bm\theta_o>\textbf{r}_o\}}(\bm\theta_o)$, $1_{\{\bm\theta_o>\textbf{r}_o\}}(\bm\theta_o)$ is the indicator function which equals 1 if $\bm\theta_o>\textbf{r}_o$ holds, and 0 otherwise, and $\text{Pr}_{c^*}(\cdot)$ denotes the prior probability of $\bm\theta_o>\textbf{r}_o$ under the completed prior under $H_c$.
\end{lemma}
\noindent \textbf{Proof:} Appendix A.

\begin{remark}
Note that in the special case where
\[
\pi_c(\bm\theta_o,\bm\phi)=\pi_u(\bm\theta_o,\bm\phi|\bm\theta_e=\textbf{r}_e) \text{Pr}_{u}(\bm\theta_o>\textbf{r}_o|\bm\theta_e=\textbf{r}_e)^{-1} 1_{\{\bm\theta_o>\textbf{r}_o\}}(\bm\theta_o),
\]
so that the completed prior under $H_{c^*}$ is equal to $\pi_u(\bm\theta_o,\bm\phi|\bm\theta_e=\textbf{r}_e)$,
then \eqref{SD3} results in the known generalization of the Savage-Dickey density ratio of the Bayes factor for an equality and order hypothesis against an unconstrained alternative,
\begin{equation}
B_{cu}=\frac{\pi_u(\bm\theta_e=\textbf{r}_e|\textbf{y})}{\pi_u(\bm\theta_e=\textbf{r}_e)}\times \frac{\text{Pr}_{u}(\bm\theta_o>\textbf{r}_o|\textbf{y},\bm\theta_e=\textbf{r}_e)}
{\text{Pr}_{u}(\bm\theta_o>\textbf{r}_o|\bm\theta_e=\textbf{r}_e)}.
\label{BFSD1}
\end{equation}
This expression has been reported in \cite{MulderGelissen:2018}, for example.
\end{remark}

\begin{remark}
In the special case with no order constraints, the parameters $\bm\theta_o$ would be part of the nuisance parameters $\bm\phi$, and thus \eqref{SD3} becomes equal to \eqref{SD2}.
\end{remark}


\begin{remark}
The importance of the ``completed'' prior where the one-sided constraints are omitted was also highlighted by \cite{Pericchi:2008} for intrinsic Bayes factors.
\end{remark}

Lemma \ref{lemmaSD} shows which four ingredients need to be computed in order to obtain the Bayes factor of a constrained hypothesis against an unconstrained alternative. The computation of these four ingredients can be done in different ways across different statistical models. To give readers more insights about the computational aspects, the next section shows the application of the result under two different statistical models: the multivariate normal model for multivariate continuous data and the multinomial model for categorical data.

\section{Applications}\label{Section3}

\subsection{A multivariate $t$ test using the JZS Bayes factor}\label{App1}
The Cauchy prior for standardized effects is becoming increasingly popular for Bayes factor testing in the social and behavioral sciences \citep{Rouder:2009,Rouder:2012,Rouder:2015}. This Bayes factor is based on key contributions by \cite{Jeffreys}, \cite{ZellnerSiow1980}, and \cite{Liang:2008}, and is therefore also referred to as the JZS Bayes factor. Here we extend this to a Bayesian multivariate $t$ test under the multivariate normal model, and show how to compute the Bayes factor for testing a hypothesis with equality and order constraints on the standardized effects using Lemma \ref{lemmaSD}. Note that this test differs from multivariate $t$ tests on multiple coefficients using a multivariate Cauchy prior under univariate linear regression models \citep{Rouder:2015,Heck:2020} as we consider a model with a multivariate outcome variable.

Let a multivariate dependent variable of $p$ dimensions, $\textbf{y}_i$, follow a multivariate normal distribution, i.e., $\textbf{y}_i\sim N(\bm\mu,\bm\Sigma)$, for $i=1,\ldots,n$. To explicitly model the standardized effects, we reparameterize the model according to
\begin{equation}
\textbf{y}_i\sim N(\textbf{L}_{\bm\Sigma}\bm\delta,\bm\Sigma),
\label{model1}
\end{equation}
where $\bm\delta$ are the unknown standardized effects, and $\textbf{L}_{\bm\Sigma}$ is the lower triangular Cholesky factor of the unknown covariance matrix $\bm\Sigma$, such that $\textbf{L}_{\bm\Sigma}\textbf{L}_{\bm\Sigma}'=\bm\Sigma$. The model in \eqref{model1} is a generalization of the univariate model considered by \cite{Rouder:2009}, $y_i\sim N(\sigma\delta,\sigma^2)$.

As a motivating example we consider the bivariate data set ($p=2$) presented in \cite{Larocque2004}, where $\textbf{y}_i=(y_{i1},y_{i2})'$ contains the cell count differences of CD45RA T and CD45RO T cells of $n=36$ HIV-positive newborn infants \citep{Sleasman}. We are interested in testing whether the standardized effects of the cell count differences of the two cell types are equal and positive, i.e.,
\begin{eqnarray*}
H_c &:& \delta_1 = \delta_2 > 0\\
H_u &:& (\delta_1,\delta_2)\in\mathbb{R}^2.
\end{eqnarray*}
The sample means were $\bar{\textbf{y}}=(86.94,193.47)'$ and the estimated covariance matrix equalled $\hat{\bm\Sigma}=[20197 ~ 23515; 23515~ 106350]$.

Extending the prior proposed by \cite{Rouder:2009} to the multivariate normal model, we set an unconstrained Cauchy prior on $\bm\delta$ under $H_u$ and the Jeffreys prior for the covariance matrix:
\begin{eqnarray*}
\pi_u(\bm\delta,\bm\Sigma) &=& \pi_u(\bm\delta)\times \pi_u(\bm\Sigma)\\
&=& \text{Cauchy}(\bm\delta|\textbf{S}_{u,0}) \times |\bm\Sigma|^{-\frac{p+1}{2}}.
\end{eqnarray*}
A diagonal prior scale matrix is set for $\delta$ given by $\textbf{S}_{u,0}=\text{diag}(s_1^2,s_2^2)$, with $s_1^2=s_2^2=.25$. This prior implies that standardized effects of about 0.5 are likely under $H_u$.
Under the constrained hypothesis $H_c$ the free parameters are the common standardized effect, say, $\delta=\delta_1=\delta_2$, and the error covariance matrix, $\bm\Sigma$. We set a univariate Cauchy prior for $\delta$ with scale $s_{1}$ truncated in $\delta>0$, and the Jeffreys prior for $\bm\Sigma$, i.e., 
\begin{eqnarray*}
\pi_c(\delta,\bm\Sigma) &=& \pi_1(\delta)\times \pi_1(\bm\Sigma)\\
&=& 2\times \text{Cauchy}(\delta|s_{1}) \times 1(\delta>0) \times |\bm\Sigma|^{-\frac{p+1}{2}},
\end{eqnarray*}
where $\pi_{c^*}(\delta)=\text{Cauchy}(\delta|s_{1})$ denotes the completed prior, and 2 serves as a normalizing constant for the completed prior as $\text{Pr}_{c^*}(\delta>0)^{-1}=2$. As $\delta$ has a similar interpretation as $\delta_1$ and $\delta_2$ under $H_u$, the prior scale is also set to $s_{1}=.5$.

By applying the following linear transformation on the standardized effects,
\begin{equation}
\bm\theta=
\left[
\begin{array}{c}
\theta_{e}\\
\theta_{o}
\end{array}
\right]
=
\left[
\begin{array}{c}
\delta_{1} - \delta_2\\
\delta_{2}
\end{array}
\right]
\left[
\begin{array}{cc}
1 & -1 \\
0 & 1
\end{array}
\right]
\left[
\begin{array}{c}
\delta_1\\
\delta_2
\end{array}
\right]= \textbf{T}\bm\delta,
\label{transformation}
\end{equation}
the model can equivalently be written as $\textbf{y}_i\sim N(\textbf{L}\textbf{T}^{-1}\bm\theta,\bm\Sigma)$, and the hypotheses can be written as
\begin{eqnarray*}
H_c &:& \theta_e=0,\theta_o>0\\
H_u &:& (\theta_e,\theta_o)\in\mathbb{R}^2.
\end{eqnarray*}
Note here that $\theta_o$ corresponds to the common standardized effect $\delta$ under $H_c$. The prior for $(\theta_e,\theta_o)$ under $H_u$ follows a bivariate Cauchy distribution with scale matrix $\textbf{T}\textbf{S}_{u,0}\textbf{T}'=[0.5~-0.25;-0.25~0.25]$.

If one would be testing the hypotheses with the Savage-Dickey density ratio in \eqref{BFSD1}, it is easy to show that the implied prior for $\delta$ under $H_c$ (i.e., the conditional unconstrained prior for $\theta_o$ given $\theta_e=0$ under $H_u$) follows a Student $t$ distribution with 2 degrees of freedom with a scale parameter of $0.25^2=0.125$; thus assuming that standardized effects of 0.25 are likely under $H_c$. As was discussed earlier, there is no logical reason why the common standardized effect under the restricted hypothesis $H_c$ is expected to be smaller than the standardized effects under $H_u$ a priori.

\begin{figure}
\includegraphics[width=13.5cm]{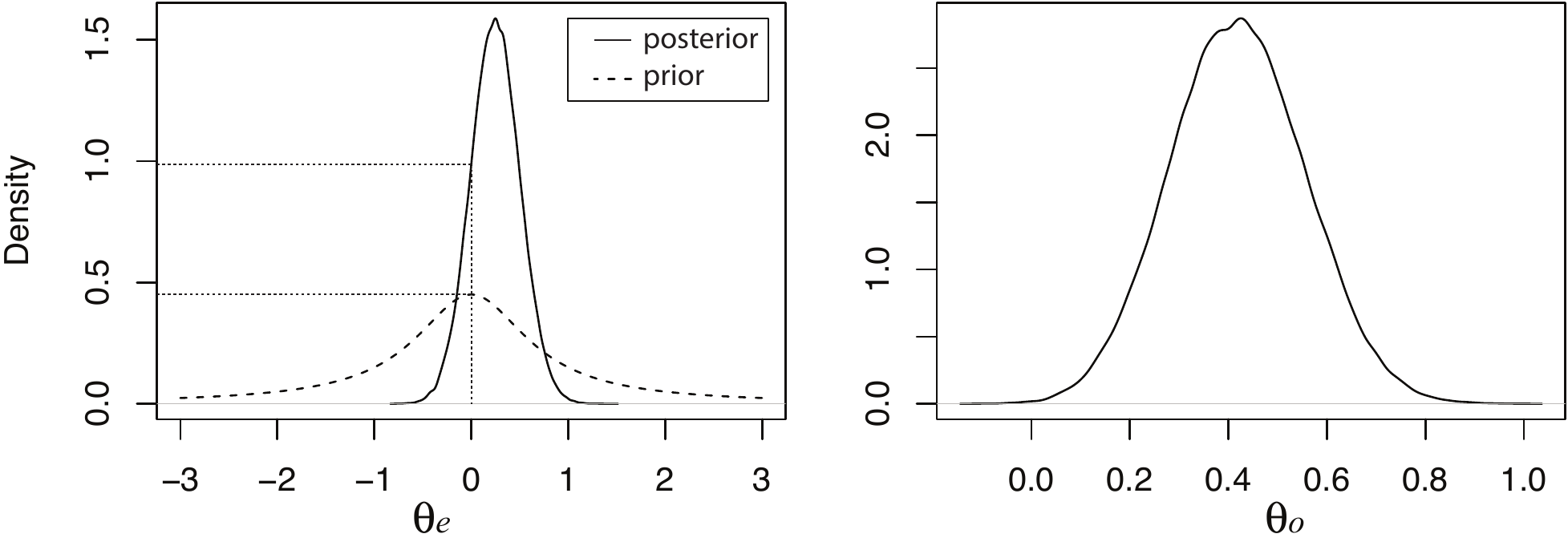}
\caption{Estimated probability densities for the multivariate Student $t$ test. Left panel. Marginal posterior (solid line) and prior (dashed line) for $\theta_e=\delta_1-\delta_2$. The dotted lines indicate the estimated density values at $\theta_e=0$. Right panel. Estimated conditional posterior for $\theta_o$ given $\theta_e=0$ under $H_u$.}
\label{densplots}
\end{figure}

The JSZ Bayes factor for this constrained testing problem using Lemma \ref{lemmaSD} based on the actual Cauchy priors for the standardized effects can be computed using MCMC output from a sampler under $H_u$, which is described in Appendix B. The R code for the computation is given in Appendix \ref{Rcode1}. The four key quantifies in \eqref{SD3} are computed as follows:
\begin{itemize}
\item As the unconstrained marginal prior for $\theta_e$ follows a Cauchy distribution with scale $\sqrt{.5}$ (Figure \ref{densplots}, left panel, dashed line), the prior density equals $\pi_u(\theta_e=0|\textbf{Y})=\sqrt{2}/\pi$.
\item The estimated marginal posterior for $\theta_e$ under $H_u$ follows from MCMC output. The estimated posterior for $\theta_e$ is plotted in Figure \ref{densplots} (left panel, solid line). This yields $\hat{\pi}_u(\theta_e=0|\textbf{Y})=0.9871618$.
\item As the completed prior for $\delta$ under $H_{c^*}$ follows a $\text{Cauchy}(0.5)$ distribution that is centered at zero, the prior probability equals $Pr_{c^*}(\delta>0)=0.5$.
\item As the priors for the covariance matrices cancel out in the fraction, the expected value can be written as $\mathbb{E}\left\{ \frac{ \text{Cauchy}(\theta_o|0.5)}{\text{Cauchy}(\theta_o|0.25)}
1_{\{\theta_o>0\}}(\theta_o)
\right\}$ under the conditional posterior for $\theta_o$ given $\theta_e=0$ under $H_u$. Appendix B also shows how to get posterior draws from $\theta_o$ under $H_u$ given $\theta_e=0$. The estimated posterior is displayed in Figure \ref{densplots} (right panel). A Monte Carlo estimate can then be used to compute the expectation, which yields 1.098799.
\end{itemize}

Application of Lemma \ref{lemmaSD} then yields a Bayes factor for $H_c$ against $H_u$ of $B_{cu} = \frac{0.9871618}{\sqrt{2}/\pi \times .5}\times 1.098799=4.8$. Thus there is 4.8 times more evidence in the data for equal and positive standardized count differences than for the unconstrained alternative hypothesis. Assuming equal prior probabilities for $H_c$ and $H_u$ this would yield posterior probabilities of $\text{Pr}(H_c|\textbf{Y})=.783$ and $\text{Pr}(H_u|\textbf{Y})=.217$. Thus there is mild evidence for $H_c$ relative to $H_u$. In order to draw clearer conclusions more data would need to be collected.

%
%

\subsection{Constrained hypothesis testing under the multinomial model}\label{App2}
When analyzing categorical data using a multinomial model, researchers are often interested in testing the relationships between the probabilities of the different cells \citep{Robertson:1978,Klugkist:2010,Heck:2019}. As an example we consider an experiment for testing the Mendelian inheritance theory discussed by \cite{Robertson:1978}. A total of 556 peas coming from crosses of plants from round yellow seeds and plants from wrinkled green seeds were divided in four categories. The cell probabilities for these categories are contained in the vector $\bm\gamma=(\gamma_1,\gamma_2,\gamma_3,\gamma_4)$, where $\gamma_1$ denotes the probability that a pea resulting from such a mating is round and yellow; $\gamma_2$ denotes the probability that it is wrinkled and yellow; $\gamma_3$ denotes the probability that it is round and green; and $\gamma_4$ denotes the probability that it is wrinkled and green. The Mendelian theory states that $\gamma_1$ is largest, followed by $\gamma_2$ and $\gamma_3$ which are assumed to be equal, and $\gamma_4$ is expected to be smallest. This can be summarized as $H_c:\gamma_1>\gamma_2=\gamma_3>\gamma_4$. In particular the theory dictates that the four probabilities are proportional to 9, 3, 3, and 1, respectively. We translate this to a completed prior under $H_{c^*}$ such that its means satisfy $\frac{E(\gamma_1)}{E(\gamma_2)}=\frac{E(\gamma_2)}{E(\gamma_4)}=3$. This can be achieved via a Dirichlet prior under an alternative parameterization, $(\xi_1,\xi_2,\xi_4)\sim \text{Dirichlet}(\alpha_{c1},\alpha_{c2},\alpha_{c3})$, with $\bm\alpha_{c}=(9,6,1)'$. The cell probabilities under $H_{c^*}$ are then defined by $(\gamma_1,\gamma_2,\gamma_4) = (\xi_1,\xi_2/2,\xi_4)$, which then follow a specific scaled Dirichlet distribution, which we denote by SDirichlet$(9,6,1)$\footnote{This specific scaled Dirichlet distribution has probability density function $\pi_{c^*}(\gamma_1,\gamma_2,\gamma_4) = \text{SDirichlet}(\alpha_{c1},\alpha_{c2},\alpha_{c3}) = \frac{2^{\alpha_{c2}}}{B(\alpha_{c1},\alpha_{c2},\alpha_{c3})} \gamma_1^{\alpha_{c1}-1} \gamma_2^{\alpha_{c2}-1} (1-\gamma_1-2\gamma_2)^{\alpha_{c3}-1}$, with $\gamma_4=1-\gamma_1-2\gamma_2$, where $B(\cdot)$ is the multivariate beta function.}. The prior for the cell probabilities under $H_c$ is then a truncation of this scaled Dirichlet distribution truncated under $\gamma_1>\gamma_2>\gamma_4$. The Mendelian hypothesis can equivalently be formulated on the transformed parameters $(\theta_e,\theta_{o,1},\theta_{o,2},\phi)=(\gamma_2-\gamma_3,\gamma_1-\gamma_2,\gamma_2-\gamma_4,\gamma_2)$ so that $H_c:\theta_e=0,(\theta_{o,1},\theta_{o,2})>\textbf{0}$, as in Lemma \ref{lemmaSD}. It is easier however to compute the four quantities in \eqref{SD3} via the untransformed parameters $\bm\gamma$ as will be shown below.

The Mendelian hypothesis will be tested against an unconstrained alternative which does not make any assumptions about the relationships between the cell probabilities. A uniform prior on the simplex will be used under the alternative, i.e., $\pi_u(\gamma_1,\gamma_2,\gamma_3,\gamma_4)=\text{Dirichlet}(1,1,1,1)$. The observed frequencies in the four respective categories were equal to 315, 101, 108, and 32.

The R code for the computation of the Bayes factor of $H_c$ against $H_u$ can be found in Appendix \ref{Rcode2}.
\begin{itemize}
\item The unconstrained marginal prior density at $\theta_e=0$ can be estimated from a sample of $\theta_e=\gamma_2-\gamma_3$ where $\bm\gamma$ is sampled from the unconstrained Dirichlet$(1,1,1,1)$ prior, resulting in $\hat{\pi}_u(\theta_e=0)=1.476556$.
\item Similarly, the unconstrained marginal posterior density at $\theta_e=0$ can be obtained by sampling $\bm\gamma$ from the unconstrained Dirichlet$(316,102,109,33)$ posterior, resulting in $\hat{\pi}_u(\theta_e=0|\textbf{y})=13.71403$.
\item The prior probability under $H_c$ can be obtained by first sampling $(\xi_1,\xi_2,\xi_4)\sim \text{Dirichlet}(9,6,1)$, then transforming the prior draws according to $(\gamma_1,\gamma_2,\gamma_4) = (\xi_1,\xi_2/2,\xi_4)$, and taking the proportion of draws satisfying the constraints $\text{Pr}_c(\gamma_1>\gamma_2>\gamma_3)\approx S^{-1}\sum_{s=1}^S I(\gamma_1^{(s)}>\gamma_2^{(s)}>\gamma_3^{(s)})=0.8949818$, where $\bm\gamma^{(s)}$ denotes the $s$-th draw, for $s=1,\ldots,S$.
\item To get draws from the conditional distribution $(\gamma_1,\gamma_2,\gamma_3,\gamma_4)$ given $\gamma_2=\gamma_3$ when $(\gamma_1,\gamma_2,\gamma_3,\gamma_4)\sim \text{Dirichlet}(\alpha_1,\alpha_2,\alpha_3,\alpha_4)$ under $H_u$, we can sample transformed parameters $(\xi_1,\xi_2,\xi_4) \sim \text{Dirichlet}(\alpha_1,\alpha_2+\alpha_3-1,\alpha_4)$, and compute $(\gamma_1,\gamma_2,\gamma_3,\gamma_4)=(\xi_1,\xi_2/2,\xi_2/2,\xi_4)$. This can be used to obtain draws from the conditional posterior for $(\gamma_1,\gamma_2,\gamma_3,\gamma_4)$ given $\gamma_2=\gamma_3$ under $H_u$ by setting $\bm\alpha=(315,101,108,33)$. The expectation in \eqref{SD3} can then be computed as the arithmetic mean of $\frac{\text{SDirichlet}((\gamma_1,\gamma_2,\gamma_4)|\bm\alpha=(9,6,1))}{\text{SDirichlet}((\gamma_1,\gamma_2,\gamma_4)|\bm\alpha=(1,1,1))}I(\gamma_1>\gamma_2>\gamma_4)$ based on a sufficiently large sample. This yields an estimate of 10.50881.
\end{itemize}

In sum the Bayes factor of the Mendelian hypothesis against the noninformative unconstrained alternative is equal to $B_{cu}=\frac{13.71403}{1.476556\times 0.8949818}\times 10.50881=109.0572$. This can be interpreted as relatively strong evidence for the Mendelian hypothesis against an unconstrained alternative based on the observed data.

Finally note that by using probability calculus it can be shown that the first two ingredients have analytic solutions as the marginal probability density at $\theta_e=\gamma_2-\gamma_3=0$ under $H_u$, when $\bm\gamma\sim\text{Dirichlet}(\bm\alpha)$, is equal to $\frac{\Gamma(\alpha_2+\alpha_3)(\alpha_1+\alpha_2+\alpha_3+\alpha_4-1)}{\Gamma(\alpha_2)\Gamma(\alpha_3)(\alpha_2+\alpha_3-1)2^{\alpha_2+\alpha_3-1}}$. In the above calculation, numerical estimates were used to give readers more insights how to obtain these quantities when analytic expressions are unavailable.

\section{Concluding remarks}
As Bayes factors are becoming increasingly popular to test hypotheses with equality as well as order constraints on the parameters of interest, more flexible and fast estimation methods to acquire these Bayes factors are needed. The generalization of the Savage-Dickey density ratio that was presented in this paper will be a useful contribution for this purpose. The expression allows one to compute Bayes factors in a straightforward manner from MCMC output while being able to freely specify the priors for the free parameters under the competing hypotheses. The applicability of the proposed methodology was illustrated in a constrained multivariate $t$ test using a novel extension of the JSZ Bayes factor to the multivariate normal model and in a constrained hypothesis test under the multinomial model.

\noindent

\section*{Acknowledgements}
The authors would like to thank Florian B\"{o}ing-Messing for helpful discussions at an early stage of the paper, and the editor and three anonymous reviewers for constructive feedback which improved the readability of the manuscript. The first author is supported by an ERC Starting Grant (758791).

\appendix
\section{Proof of Lemma \ref{lemmaSD}}
As the constrained model $H_c:\bm\theta_e=\textbf{r}_e~\&~\bm\theta_o>\textbf{r}_o$ is nested in the unconstrained model $H_u$, the likelihood under $H_c$ can be written as the truncation of the unconstrained likelihood, i.e., $p_c(\textbf{y}|\bm\theta_o,\bm\phi)=p_u(\textbf{y}|\bm\theta_e=\textbf{r}_e,\bm\theta_o,\bm\phi)1_{\{\bm\theta_o>\textbf{r}_o\}}(\bm\theta_o)$. The result in Lemma \ref{lemmaSD} then follows via the following steps,
\begin{eqnarray*}
B_{cu} &=& \frac{p_c(\textbf{y})}{p_u(\textbf{y})} = \frac{\iint_{\bm\theta_o>\textbf{r}_o}p_c(\textbf{y}|\bm\theta_o,\bm\phi)\pi_c(\bm\theta_o,\bm\phi)d\bm\theta_o d\bm\phi}
{\iiint p_u(\textbf{y}|\bm\theta_e,\bm\theta_o,\bm\phi)\pi_u(\bm\theta_e,\bm\theta_o,\bm\phi)d\bm\theta_e d\bm\theta_o d\bm\phi}\\
&=& \iint_{\bm\theta_o>\textbf{r}_o}\frac{p_u(\textbf{y}|\bm\theta_e=\textbf{r}_e,\bm\theta_o,\bm\phi)1_{\{\bm\theta_o>\textbf{r}_o\}}(\bm\theta_o)\pi_c(\bm\theta_o,\bm\phi)}
{p_u(\textbf{y})\pi_u(\bm\theta_e=\textbf{r}_e|\textbf{y})}d\bm\theta_o d\bm\phi \\
&& \times \pi_u(\bm\theta_e=\textbf{r}_e|\textbf{y})\\
&=& \iint_{\bm\theta_o>\textbf{r}_o}\frac{p_u(\textbf{y}|\bm\theta_e=\textbf{r}_e,\bm\theta_o,\bm\phi)\pi_c(\bm\theta_o,\bm\phi)}
{p_u(\textbf{y})\pi_u(\bm\theta_e=\textbf{r}_e,\bm\theta_o,\bm\phi|\textbf{y})}
\pi_u(\bm\theta_o,\bm\phi|\textbf{y},\bm\theta_e=\textbf{r}_e) d\bm\theta_o d\bm\phi \\
&& \times \pi_u(\bm\theta_e=\textbf{r}_e|\textbf{y})\\
&=& \iint_{\bm\theta_o>\textbf{r}_o} \frac{\pi_c(\bm\theta_o,\bm\phi)}
{\pi_u(\bm\theta_e=\textbf{r}_e,\bm\theta_o,\bm\phi)}
\pi_u(\bm\theta_o,\bm\phi|\textbf{y},\bm\theta_e=\textbf{r}_e) d\bm\theta_o d\bm\phi \\
&& \times \pi_u(\bm\theta_e=\textbf{r}_e|\textbf{y})\\
&=& \iint_{\bm\theta_o>\textbf{r}_o} \frac{\pi_c(\bm\theta_o,\bm\phi)}
{\pi_u(\bm\theta_o,\bm\phi|\bm\theta_e=\textbf{r}_e)}
\pi_u(\bm\theta_o,\bm\phi|\textbf{y},\bm\theta_e=\textbf{r}_e) d\bm\theta_o d\bm\phi \\
&& \times \frac{\pi_u(\bm\theta_e=\textbf{r}_e|\textbf{y})}{\pi_u(\bm\theta_e=\textbf{r}_e)}\\
&=& \iint \frac{\pi_{c^*}(\bm\theta_o,\bm\phi)1_{\{\bm\theta_o>\textbf{r}_o\}}(\bm\theta_o)}
{\pi_u(\bm\theta_o,\bm\phi|\bm\theta_e=\textbf{r}_e)\text{Pr}_{c^*}(\bm\theta_o>\textbf{r}_o)}
\pi_u(\bm\theta_o,\bm\phi|\textbf{y},\bm\theta_e=\textbf{r}_e) d\bm\theta_o d\bm\phi \\
&& \times \frac{\pi_u(\bm\theta_e=\textbf{r}_e|\textbf{y})}{\pi_u(\bm\theta_e=\textbf{r}_e)}\\
&=& \iint \frac{\pi_{c^*}(\bm\theta_o,\bm\phi)1_{\{\bm\theta_o>\textbf{r}_o\}}(\bm\theta_o)}
{\pi_u(\bm\theta_o,\bm\phi|\bm\theta_e=\textbf{r}_e)}
\pi_u(\bm\theta_o,\bm\phi|\textbf{y},\bm\theta_e=\textbf{r}_e) d\bm\theta_o d\bm\phi \\
&& \times \text{Pr}_{c^*}(\bm\theta_o>\textbf{r}_o)^{-1}\times \frac{\pi_u(\bm\theta_e=\textbf{r}_e|\textbf{y})}{\pi_u(\bm\theta_e=\textbf{r}_e)},
\end{eqnarray*}
which completes the proof. 
Note that in the third step the indicator function, $1_{\{\bm\theta_o>\textbf{r}_o\}}(\bm\theta_o)$, was omitted as the integrand is integrated over the subspace where $\bm\theta_o>\textbf{r}_o$. In the second last step, the completed version of the constrained hypothesis has the order constraints omitted, i.e., $H_{c^*}:\bm\theta_e=\textbf{r}_e$, with completed prior $\pi_{c^*}(\bm\theta_o,\bm\phi)$, such that $\pi_{c}(\bm\theta_o,\bm\phi)=\pi_{c^*}(\bm\theta_o,\bm\phi)\text{Pr}_{c^*}(\bm\theta_o>\textbf{r}_o)^{-1}1_{\{\bm\theta_o>\textbf{r}_o\}}(\bm\theta_o)$.

\section{MCMC sampler for the multivariate Student $t$ test}
\begin{enumerate}
\item \underline{Drawing the standardized effects $\bm\delta$.}
It is well-known that a multivariate Cauchy prior of $p$ dimensions can be written as a Multivariate normal distribution with an inverse Wishart mixing distribution on the normal covariance matrix with $p$ degrees of freedom, i.e.,
\begin{eqnarray*}
\pi_u(\bm\delta) &=& \text{Cauchy}(\bm\delta|\textbf{S}_0)\\
&=& \int N(\bm\delta|\textbf{0},\bm\Phi) \times IW(\bm\Phi|p,\textbf{S}_0) d\bm\Phi.
\end{eqnarray*}
Thus the conditional prior for $\bm\delta$ given the auxiliary parameter matrix $\bm\Phi$ follows a $N(\textbf{0},\bm\Phi)$ distribution. Consequently, as $\textbf{z}_{\bm\Sigma,i}=\textbf{L}_{\bm\Sigma}^{-1}\textbf{y}_i\sim N(\bm\delta,\textbf{I}_p)$, the conditional posterior of $\bm\delta$ follows a multivariate normal posterior,
\[
\bm\delta | \bm\Phi, \bm\Sigma,\textbf{y} \sim N(n(\bm\Phi^{-1} + n\textbf{I}_p)^{-1}\bar{\textbf{z}}_{\bm\Sigma},
(\bm\Phi^{-1} + n\textbf{I}_p)^{-1}),
\]
where $\bar{\textbf{z}}_{\bm\Sigma}$ are the sample means of $\textbf{z}_{\bm\Sigma,i}$, for $i=1,\ldots,n$.
\item \underline{Drawing the auxiliary covariance matrix $\bm\Phi$.}
The conditional posterior for $\bm\Phi$ only depends on the standardized effects and it follows an inverse Wishart distribution,
\[
\bm\Phi |\bm\delta \sim IW(p+1,\textbf{S}_0+\bm\delta\bm\delta').
\]
\item \underline{Drawing the error covariance matrix $\bm\Sigma$.}
The conditional posterior for the covariance matrix does not follow a known distribution. For this reason we use a random walk \citep[e.g.,][]{Gelman} for sampling the separate elements of $\bm\Sigma$.
\end{enumerate}

The sampler under the unconstrained model while restricting $\delta_1=\delta_2$ ($=\delta$) is very similar except that the prior for $\delta$ is now univariate Cauchy$(\delta|0.25)$ and $\bm\Phi=[\phi^2]$ is a scalar, and thus the conditional posterior for $\delta$ is univariate normal $N(2n(\phi^{-2} + 2n)^{-1}\bar{z}_{\bm\Sigma},(\phi^{-2} + n)^{-1})$, where $\bar{z}_{\bm\Sigma}$ is the mean of $\bar{\textbf{z}}_{\bm\Sigma}$. Also note that the inverse Wishart distribution in Step 2 is now for a $1\times 1$ covariance matrix which is equivalent to an inverse gamma distribution.


\section{R code for empirical analyses}
\subsection{R code for multivariate $t$ test in Section \ref{App1}}\label{Rcode1}

\begin{verbatim}
library(mvtnorm)
library(Matrix)

# computing the unconstrained marginal prior density at \theta_e=0:
priorE <- dcauchy(0, location = 0, scale = sqrt(.5))

# computing the unconstrained marginal posterior density at \theta_e=0:
# read data
Y <- t(matrix(c(242,1708,569,569,270,757,-25,499,309,231,22,338,-42,26,
  -233,119,206,163,-106,-186,55,54,85,48,30,50,194,525,-87,-110,159,148,
  29,102,89,364,-9,36,158,234,76,122,15,24,3,36,93,71,160,44,66,128,180,
  155,237,85,105,76,16,6,167,364,-10,-18,-61,-21,-7,-2,15,32,160,188),
  nrow=2))
set.seed(123)
#dimension
p <- ncol(Y)
nums <- p*(p+1)/2
n <- nrow(Y)
#initial parameter values based on burn-in period
delta <- c(.5,.2)
Sigma <- matrix(c(2,2,2,11),2,2) * 10**4
L <- t(chol(Sigma))
Phi <- diag(p)
#selection of unique elements in \Sigma
lowerSigma <- lower.tri(Sigma,diag=TRUE)
welklower <- which(lowerSigma)
# tranformation matrix
Trans <- matrix(c(1,0,-1,1),ncol=2)
#prior hyperparameters
S0 <- diag(p) * .5**2
# random walk sd's for the elements of \Sigma to have an
# efficient acceptance probability based on burn-in period.
sdstep <- c(9,13,48) * 10**3
#store draws
numdraws <- 1e5
storeDelta <- matrix(0,nrow=numdraws,ncol=p)
storeSigma <- storePhi <- array(0,dim=c(numdraws,p,p))
#draws from stationary distribution
for(s in 1:numdraws){
  #draw delta
  deltaMean <- c(apply(Y%*%t(solve(L)),2,mean))
  SigmaDelta <- solve(n*diag(p) + solve(Phi))
  muDelta <- c(SigmaDelta%*%deltaMean*n)
  delta <- c(rmvnorm(1,mean=muDelta,sigma=SigmaDelta))

  #draw Phi
  Phi <- solve(rWishart(1,df=p+1,Sigma=solve(S0 + delta%*%t(delta)))[,,1])
  #draw Sigma using MH

  for(sig in 1:nums){
    welknu <- welklower[sig]
    step1 <- rnorm(1,sd=sdstep[sig])
    Sigma0 <- matrix(0,p,p)
    Sigma0[lowerSigma] <- Sigma[lowerSigma]
    Sigma0[welknu] <- Sigma0[welknu] + step1
    Sigma_can <- Sigma0 + t(Sigma0) - diag(diag(Sigma0))

    if(min(eigen(Sigma_can)$values) > .000001){
      #the candidate is positive definite
      L_can <- t(chol(Sigma_can))
      #acceptance probability
      R_MH <- exp( sum(dmvnorm(Y,mean=c(L_can%*%delta),sigma=Sigma_can,
        log=TRUE)) - (p+1)/2*log(det(Sigma_can)) -
        sum(dmvnorm(Y,mean=c(L%*%delta),sigma=Sigma,log=TRUE)) +
        (p+1)/2*log(det(Sigma)) )
      if(runif(1) < R_MH){
        #accept draw
        Sigma <- Sigma_can
        L <- t(chol(Sigma))
      }
    }
  }

  storeDelta[s,] <- delta
  storeSigma[s,,] <- Sigma
  storePhi[s,,] <- Phi
}
drawsE <- storeDelta[,1] - storeDelta[,2]
denspost <- density(drawsE)
df <- approxfun(denspost)
postE <- df(0)
# Figure 1 (left panel)
plot(denspost,xlim=c(-3,3),main="",xlab="theta_e")
seq1 <- seq(-3,3,length=1e3)
lines(seq1,dcauchy(seq1,scale=sqrt(.5)),lty=2)

# computing the prior probability of \theta_o>0 under H_c:
priorO <- 1 - pcauchy(0, location = 0, scale = .5)

# computing the expectation of the ratio of the priors
# from a posterior sample under H_c given \theta_e = 0
# initialization
set.seed(123)
p1 <- 1
p <- ncol(Y)
nums <- p*(p+1)/2
n <- nrow(Y)
S0 <- diag(1)*.25**2
# initial parameter values based on burn-in period
delta <- .55
Phi <- matrix(1)
Sigma <- matrix(c(23,22,22,89),nrow=2) * 10**3
L <- t(chol(Sigma))
# random walk sd's for the elements of \Sigma to have an
# efficient acceptance probability based on burn-in period.
sdstep1 <- c(10,15,48) * 10**3
lowerSigma <- lower.tri(Sigma,diag=TRUE)
welklower <- which(lowerSigma)
# store draws
numdraws <- 1e5
storeDelta1 <- matrix(0,nrow=numdraws,ncol=1)
storePhi1 <- array(0,dim=c(numdraws,p1,p1))
storeSigma1 <- array(0,dim=c(numdraws,p,p))
for(s in 1:numdraws){
  #draw delta
  deltaMean <- mean(c(apply(Y%*%t(solve(L)),2,mean)))
  SigmaDelta <- solve(2*n*diag(p1) + solve(Phi))
  muDelta <- c(SigmaDelta%*%deltaMean*2*n)
  delta <- c(rmvnorm(1,mean=muDelta,sigma=SigmaDelta))

  #draw Phi
  Phi <- solve(rWishart(1,df=p1+1,Sigma=solve(S0 +
    delta%*%t(delta)))[,,1])

  #draw Sigma using MH
  deltavec <- rep(delta,2)
  for(sig in 1:nums){
    welknu <- welklower[sig]
    step1 <- rnorm(1,sd=sdstep1[sig])
    Sigma0 <- matrix(0,p,p)
    Sigma0[lowerSigma] <- Sigma0[lowerSigma] + Sigma[lowerSigma]
    Sigma0[welknu] <- Sigma0[welknu] + step1
    Sigma_can <- Sigma0 + t(Sigma0) - diag(diag(Sigma0))

    if(min(eigen(Sigma_can)$values) > .000001 ){
      #the candidate is positive definite
      L_can <- t(chol(Sigma_can))
      #dit zou sneller kunnen via onafhankelijke univariate normals
      R_MH <- exp( sum(dmvnorm(Y,mean=c(L_can%*%deltavec),
        sigma=Sigma_can,log=TRUE)) - (p+1)/2*log(det(Sigma_can)) -
        sum(dmvnorm(Y,mean=c(L%*%deltavec),sigma=Sigma,log=TRUE))
        + (p+1)/2*log(det(Sigma)) )
      if(runif(1) < R_MH){
        #accept draw
        Sigma <- Sigma_can
        L <- t(chol(Sigma))
      }
    }
  }

  storeDelta1[s,] <- delta
  storePhi1[s,,] <- Phi
  storeSigma1[s,,] <- Sigma
}
expratio <- mean(dcauchy(c(storeDelta1),scale=.5) /
  dcauchy(c(storeDelta1),scale=.25)
  * (c(storeDelta1)>0))
# Figure 1, right panel
plot(density(c(storeDelta1)),main="",xlab="theta_o")

# computation of the Bayes factor
Bcu <- postE / (priorE * priorO) * expratio
\end{verbatim}

\subsection{R code for multinomial model in Section \ref{App2}}\label{Rcode2}
\begin{verbatim}
library(MCMCpack)
set.seed(123)

# computing the unconstrained marginal prior density at \theta_e=0:
uncpriorsample <- rdirichlet(n=1e7, alpha=c(1,1,1,1))
densprior <- density(uncpriorsample[,2]-uncpriorsample[,3])
df <- approxfun(densprior)
priorE <- df(0)
remove(uncpriorsample)

# computing the unconstrained marginal posterior density at \theta_e=0:
uncpostsample <- rdirichlet(n=1e7, alpha=c(1+315,1+101,1+108,1+32))
denspost <- density(uncpostsample[,2]-uncpostsample[,3])
df <- approxfun(denspost)
postE <- df(0)
remove(uncpostsample)

# computing the prior probability of \theta_o>0 under H_c:
priorsample1 <- rdirichlet(n=1e7,alph=c(9,6,1))
priorsample1[,2] <- priorsample1[,2]/2
priorO <- mean(priorsample1[,1] > priorsample1[,2] &
                 priorsample1[,2] > priorsample1[,3])
remove(priorsample1)

# computing the expectation of the ratio of priors:
# first define probability density for (gamma1,gamma2)
SDirichlet <- function(gamma1,gamma2,alpha1,alpha2,alpha3){
  alphavec <- c(alpha1,alpha2,alpha3)
  B1 <- exp(sum(lgamma(alphavec)) - lgamma(sum(alphavec)))
  return(
    2^alpha2 / B1 * gamma1^(alpha1-1) * gamma2^(alpha2-1) * 
      (1-gamma1-2*gamma2)^(alpha3-1)
  )
}
condpostsample <- rdirichlet(n=1e7, alpha=c(316,210,33))
condpostsample[,2] <- condpostsample[,2]/2
expratio <- mean(SDirichlet(condpostsample[,1],condpostsample[,2],9,6,1) /
  SDirichlet(condpostsample[,1],condpostsample[,2],1,1,1) *
    (condpostsample[,1]>condpostsample[,2] & 
      condpostsample[,2]>condpostsample[,3])
  )
remove(condpostsample)

# computing the Bayes factor of $H_c$ against $H_u$:
Bcu <- postE/(priorE*priorO)*expratio
\end{verbatim}

\bibliographystyle{apacite}
\bibliography{refs_mulder}

\end{document}